\documentclass[prl,twocolumn,showpacs,superscriptaddress]{revtex4-1}

\usepackage{amssymb,amsfonts,amsmath} 
\usepackage{graphicx}
\usepackage{bm}
\usepackage{color}
\usepackage{hyperref}

\newcommand{\tn}{\textnormal} 
 
\newcommand{\I}{\mathrm{i}}
\newcommand{\D}{\mathrm{d}} 
\newcommand{\E}{\mathrm{e}}
\renewcommand{\Im}{\mathrm{Im}}

\begin{document} \title{Understanding the Josephson current through a Kondo-correlated quantum dot}

\author{D. J. Luitz} 
\affiliation{\mbox{Institut f\"ur Theoretische Physik und Astrophysik, Universit\"at W\"urzburg,
97074 W\"urzburg, Germany}}

\author{F. F. Assaad} 
\affiliation{\mbox{Institut f\"ur Theoretische Physik und Astrophysik, Universit\"at W\"urzburg,
97074 W\"urzburg, Germany}}

\author{T. Novotn{\'y}} 
\affiliation{Department of Condensed Matter Physics, Faculty of Mathematics
and Physics, Charles University, Ke Karlovu 5, 12116 Prague, Czech Republic}

\author{C. Karrasch} 
\affiliation{Department of Physics, University of California, Berkeley, California 95720, USA}

\author{V. Meden} 
\email{meden@physik.rwth-aachen.de} 
\affiliation{Institut f{\"u}r Theorie der Statistischen Physik, RWTH Aachen University and JARA---Fundamentals of Future Information
Technology, 52056 Aachen, Germany}

\begin{abstract} 

We study the Josephson current $0$-$\pi$ transition of a 
quantum dot tuned to the Kondo regime. The physics can be quantitatively 
captured by the numerically exact continuous time quantum Monte Carlo method applied 
to the single-impurity Anderson model with BCS superconducting leads. For a 
comparison to an experiment the tunnel couplings are determined by 
fitting the normal-state linear conductance.  Excellent agreement for the 
dependence of the critical Josephson current on the level energy is achieved.
For increased tunnel couplings the Kondo scale becomes comparable to the 
superconducting gap and the regime of the strongest competition between superconductivity 
and Kondo correlations is reached; we predict the gate voltage dependence of the 
critical current in this regime. 

\end{abstract}

\date{January 26, 2012} 
\maketitle

\textit{Introduction.} Recently, hybrid superconductor-quan\-tum dot devices have attracted much attention 
\cite{De-Franceschi10} due to their peculiar physical behavior determined by the interplay 
of superconductivity of the leads and the level characteristics of 
the dot. Applications in nanoelectronics or quantum-information 
processing are envisaged. Among other properties, DC Josephson transport 
\cite{Kasumov99,vanDam06,Cleuziou06,Jorgensen07,Eichler09,Maurand11} was intensively studied.
Similar to the Josephson effect of ordinary tunnel junctions \cite{Tinkham96} 
a difference $\phi \neq 0,\pi$ of the order parameter phases of the two superconductors with gap
$\Delta$ leads 
to an equilibrium Josephson current $J$ running through 
the system \cite{Kasumov99,vanDam06,Cleuziou06,Jorgensen07,Eichler09,Maurand11}. 
The focus was on carbon nanotube dots \cite{Kasumov99,Cleuziou06,Jorgensen07,Eichler09,Maurand11}
with well separated single-particle levels, i.e., level broadening $\Gamma$ and temperature $T$ 
much smaller than the level spacings, simplifying the modeling as a single-level 
dot with energy $\epsilon$ can be considered. %

It is well established both theoretically \cite{Glazman89} as well as experimentally
\cite{vanDam06,Cleuziou06,Jorgensen07,Eichler09,Maurand11} that the local Coulomb 
interaction, i.e.~the dot charging energy $U$, can lead to a $0$-$\pi$ transition 
of the quantum dot Josephson junction, associated to a first order (level-crossing) 
quantum phase transition from a singlet ($0$) to a doublet ($\pi$) ground state \cite{Matsuura77}.
In fact, a variation of any of the system parameters $U$, $\epsilon$, $\Delta$, $\phi$, as well as 
the tunnel couplings $\Gamma_{L/R}$ (with $\Gamma=\Gamma_{L}+\Gamma_{R}$) can be 
used to tune the system across the phase boundary, if the others are taken from appropriate 
ranges. At $T=0$ the transition leads to a jump in $J$ from a large and positive 
($0$-phase) to a small and negative value ($\pi$-phase). At finite temperatures, it 
is smeared out and significantly diminished, yet the sign change of $J$
is clearly observed in SQUID setups %
\cite{vanDam06,Cleuziou06,Maurand11}. The experimental challenge in observing the 
true magnitude of the Josephson current to be compared with theoretical predictions consists in 
suppressing uncontrolled phase fluctuations, which can be achieved by using designed on-chip 
circuits \cite{Jorgensen07,Eichler09}. In such experiments $J$ is  
tuned by a variation of a gate voltage $V_{\rm g}$ which translates into a rather 
controlled change of $\epsilon$ \footnote{The gate 
voltage variation may simultaneously slightly change the level broadening $\Gamma$ by changing 
Schottky barriers at the contacts \cite{Maurand11}.}.

The physics becomes particularly interesting if the dot is tuned to a parameter regime
in which Kondo correlations \cite{Hewson93} become relevant for suppressed superconductivity. %
It is characterized by the appearance of the Kondo scale (at odd dot filling) 
$k_{\rm B}T_{\rm K}=\sqrt{\Gamma U/2} \exp(-\pi U/ 8 \Gamma)$  \cite{Hewson93}. 
Kondo physics is important if $k_{\rm B} T \lesssim k_{\rm B} T_{\rm K} \ll \Gamma$,
with $k_{\rm B}$ denoting the Boltzmann constant. In this regime 
perturbative methods in either $U$, such as self-consistent Hartree-Fock (HF) \cite{Shiba73}, 
or $\Gamma$ \cite{Glazman89} become uncontrolled. Even for $\Delta \gg k_{\rm B} T_{\rm K}$, 
at which superconductivity prevails, one expects Kondo correlations to have a significant 
impact on $J$. These were partly incorporated using a method 
developed for large $\Delta$ \cite{Meng09}. Other techniques successfully used for 
Kondo correlated quantum dots with normal leads, 
such as the noncrossing approximation (NCA) \cite{Clerk00,*Sellier05}, numerical 
renormalization group (NRG) \cite{Choi04,Karrasch08}, (Hirsch-Fye) quantum Monte Carlo (QMC) 
\cite{Siano04,*Siano05err}, and functional renormalization group (fRG) \cite{Karrasch08} 
were extended to the present setup. With superconducting leads they suffer from 
significant conceptual or practical limitations such as, e.g.~half filling of the 
dot level (NRG) and high (NCA, QMC) or zero (fRG) temperature and, therefore, 
cannot be used for a quantitative comparison to experiments 
performed at temperatures of the order of a few tens of mK and with a wide span of gate voltages %
\cite{Jorgensen07,Eichler09}. The regime of the strongest competition between superconductivity and Kondo correlations is reached %
for $\Delta \approx k_{\rm B} T_{\rm K}$. For typical experimental gap sizes of 
$\Delta \approx 0.1\,$meV \cite{Jorgensen07,Eichler09,Maurand11}, in this regime
$k_{\rm B} T_{\rm K} \ll \Gamma$ is no longer fulfilled. Still, even for  
$k_{\rm B} T_{\rm K} \lesssim \Gamma$ a precursor of Kondo correlations is expected %
to stabilize the singlet phase and perturbative methods become unreliable.        

\begin{figure} 
\includegraphics[width=\columnwidth]{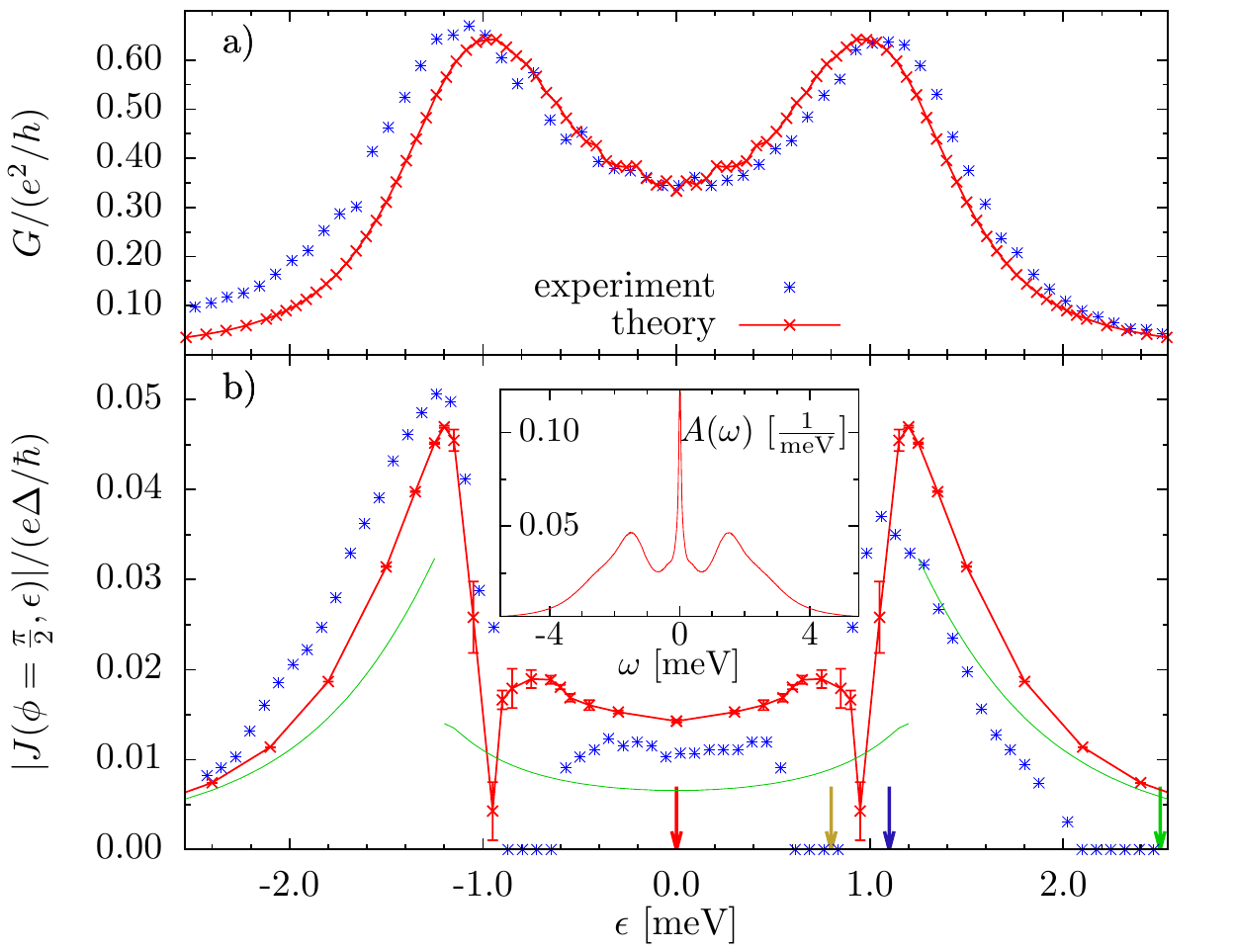} \caption{ \label{fig1} (Color online)
Comparison of experimental data~\cite{Jorgensen07} with the numerically exact solution of the superconducting Anderson model. 
a) Best fit of the normal-state linear conductance with applied magnetic field used for extracting
values of $\Gamma$ and $\Gamma_{L}/\Gamma_{R}$ (for details see the main text). b) Measured critical
current vs.~theoretically calculated Josephson current at $\phi=\frac \pi 2$ (CT-INT: symbols with line; self-consistent Hartree-Fock: thin lines). The arrows indicate the level positions for which the current phase relation is presented in Fig.~\ref{fig2}. 
{\it Inset:} Normal-state spectral function at $\epsilon=0$.} %
\end{figure}

\begin{figure} 
\includegraphics[ width=\columnwidth]{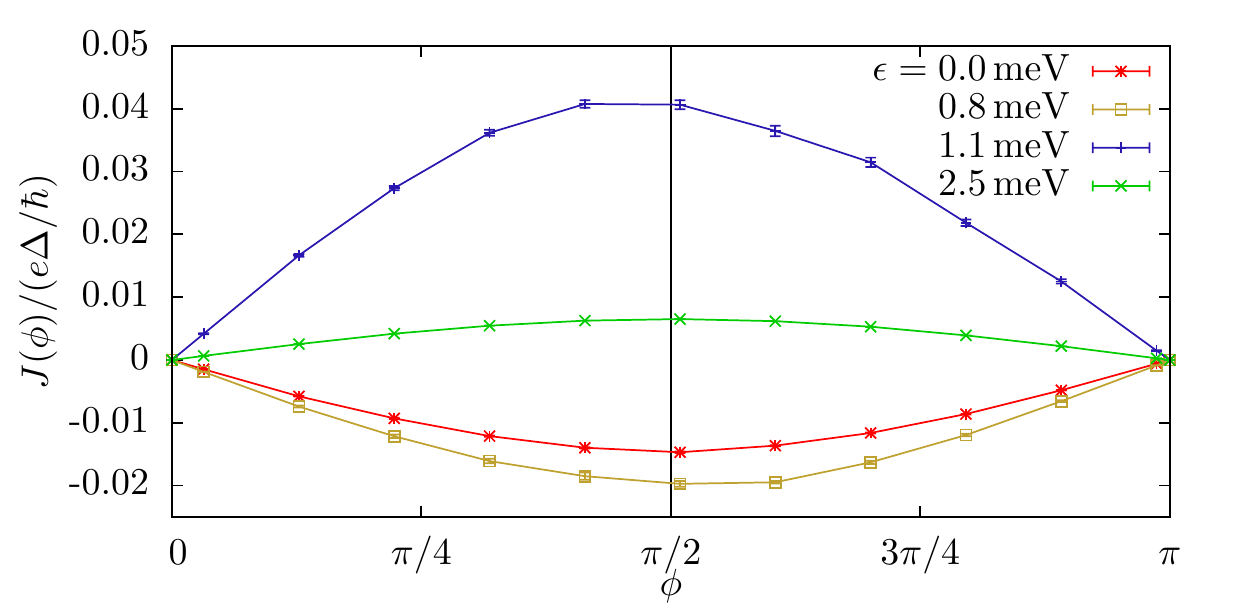} 
\caption{ \label{fig2} (Color online)
Josephson current-phase relation for the parameters of the experiment~\cite{Jorgensen07} at values of $\epsilon$ indicated by arrows in Fig.~\ref{fig1}. It is rather 
sinusoidal even very close to the critical value of $\epsilon$ ($1.1\,$meV and $0.8\,$meV) 
and the critical current is thus well approximated by $J(\phi=\frac \pi 2)$.} 
\end{figure}

\begin{figure} 
\includegraphics[ width=\columnwidth]{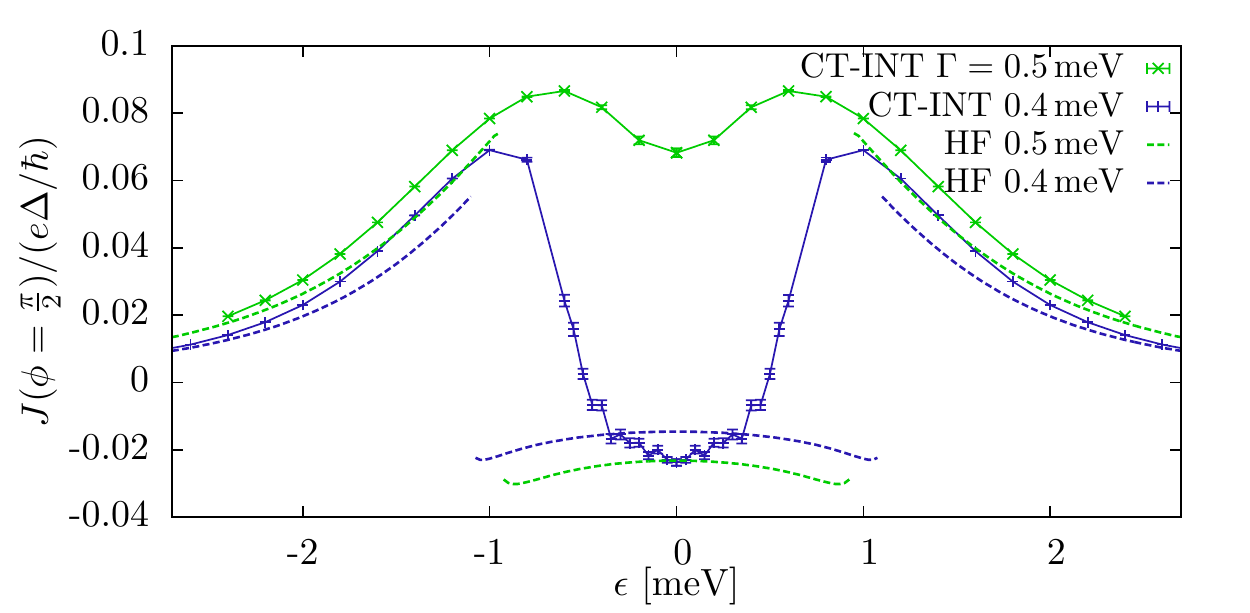} 
\caption{ \label{fig3} (Color online)
Josephson current at $\phi=\frac \pi 2$ for the parameters of the experiment  
(see Fig.~\ref{fig1}) but with increased level broadening $\Gamma=0.4\,$meV and $0.5\,$meV 
(instead of $0.27\,$meV) and thus increased $T_{\rm K}$. Self-consistent Hartree-Fock 
is obviously unable to describe the strong competition between superconductivity and Kondo 
correlations in this parameter regime and leads to a spurious $\pi$-phase for parameters 
for which the numerical exact solution only shows a precursor close to half dot filling 
$\epsilon=0$.}  
\end{figure}

Recently, the continuous time QMC method, called CT-INT 
in what follows, was introduced as a new tool to study correlated quantum dots with BCS 
leads \cite{Luitz10}. Here, we exploit the exceptional flexibility and accuracy of this 
approach and compute $J$ as well as the normal-state linear conductance $G$ for the 
parameters of the experiment of Ref.~\cite{Jorgensen07}. Our simultaneous analysis 
of $J$ and $G$ reveals that the dot shows significant Kondo correlations, 
but superconductivity prevails as $\Delta \approx 10 T_{\rm K}$. 
In the normal state it lies in the interesting and theoretically 
challenging parameter regime with $k_{\rm B} T \approx k_{\rm B} T_{\rm K} \approx \mu_{\rm B} h$, 
where $\mu_{\rm B} h$  (with the Bohr magneton $\mu_{\rm B}$) denotes the scale associated to the applied Zeeman field $h$ used to destroy %
superconductivity. 
Compared to previous approaches, we are now able to quantitatively study this experimentally
relevant  parameter regime with a numerically exact method, and
 find excellent agreement 
between the experimentally measured critical current $J_{\rm c}$  and the numerically computed one for both the $0$- 
and $\pi$-phases (see Fig.\ \ref{fig1}). We show that due to the fairly %
large left-right asymmetry of the tunnel couplings and the finite temperature the 
current-phase relation $J(\phi)$ is rather sinusoidal even close to the $0$-$\pi$ 
transition (see Fig.\ \ref{fig2}), providing an a-posteriori justification of the 
extraction of $J_{\rm c}$  from the measured current-voltage characteristics of the 
on-chip circuits applying the 
extended RSJ model \cite{Jorgensen07,Eichler09}. Finally, using the parameters of the 
experiment, but increasing $\Gamma$ such that $\Delta \approx k_{\rm B} T_{\rm K}$ we 
compute the gate voltage dependence of the current in the regime of the strongest competition between 
superconductivity and (precursors of) Kondo correlations (see Fig.~\ref{fig3}). 

\textit{Model and method.} For the description of the single level quantum dot with superconducting
leads we use the Anderson impurity model with Hamiltonian $H = H^\tn{dot} + \sum_{s=L,R}
H^\tn{lead}_s + \sum_{s=L,R}H^\tn{coup}_s$.  The dot part reads 
\begin{equation}\begin{split}
H^\tn{dot} = \sum_\sigma \epsilon_\sigma d^\dagger_\sigma d_\sigma + U \left(d^\dagger_\uparrow
d_\uparrow-\frac{1}{2}\right)\left(d^\dagger_\downarrow d_\downarrow -\frac{1}{2}\right),
\end{split}\end{equation} 
in standard second quantized notation.  In the presence
of a Zeeman field $h$ the 
single-particle energies depend on the orientation of the spin $\epsilon_\sigma=\epsilon+g \mu_{\rm
B} h \sigma /2$, 
with the Land\'e g-factor $g=2$ \cite{Tans97} and $\sigma=\pm1$. 
The energy is shifted such
that for $h=0$, $\epsilon=0$ corresponds to the point of half-filling of the dot.  The
left ($s=L$) and right ($s=R$) superconducting leads are modeled by BCS Hamiltonians
\begin{equation} H^\tn{lead}_s = \sum_{k\sigma}\epsilon_{sk}c^\dagger_{sk\sigma}c_{sk\sigma} -
\Delta \sum_k\left(\E^{\I\phi_s}c^\dagger_{sk\uparrow}c^\dagger_{s-k\downarrow} + \tn{H.c.}\right),
\end{equation} 
where (without loss of generality) $\phi_L=-\phi_R=\phi/2$.  The quantum dot is
coupled to the leads by $H^\tn{coup}_s = \sum_{k,\sigma}\left(t_{sk}
c^\dagger_{sk\sigma}d_\sigma+\tn{H.c.}\right)$.  We assume energy independent dot-lead
hybridizations $\Gamma_s=\pi \sum_{k} |t_{sk}|^2\delta(\epsilon_\text{F}-\epsilon_{sk})$, with the
Fermi energy $\epsilon_{\rm F}$.
  
The CT-INT is based on an interaction expansion of the partition function
 in which \emph{all} diagrams 
 are summed up stochastically. The method is numerically 
exact and allows the calculation of thermodynamic observables with any required precision 
$\sigma_{\text{MC}}$ (indicated by errorbars in the figures) with the practical limitation that the
computing time grows as $1/\sigma_{\text{MC}}^2$.  
Details can be found in Ref.~\cite{Luitz10}. Here we go far beyond  the proof-of-principle study of Ref.~~\cite{Luitz10} by 
considering $\epsilon \neq 0$, larger $U/\Gamma$ as well as left-right coupling asymmetries.
Furthermore, we compute the normal-state linear conductance in the challenging regime 
$k_{\rm B} T \approx k_{\rm B} T_{\rm K} \approx \mu_{\rm B} h$. 

The Josephson current is computed as the expectation value of the left (or right) current operator
$J = \I e/\hbar \sum_{k\sigma} \left< t_{Lk} c^\dagger_{Lk\sigma}d_\sigma-  \tn{H.c.} \right>$.
The noninteracting lead degrees of freedom are integrated out and 
one arrives at a formula for the Josephson current in terms of the imaginary-frequency
Nambu-Green function ${\mathbf {\mathcal G}}(\mathrm{i}\omega_n)$ of the dot only 
(directly accessible in CT-INT) \cite{Luitz10}  
\begin{equation} 
J= 2\Im \mathrm{Tr} \left[ \frac{1}{\beta} \sum_{\mathrm{i} \omega_n} \frac{\Gamma_L}{\sqrt{\Delta^2 +
\omega_n^2} } \begin{pmatrix} \mathrm{i} \omega_n & -\Delta \E^{-\I \frac{\phi}{2}} \\ \Delta \E^{\I
\frac{\phi}{2}} & -\I \omega_n  \end{pmatrix}  {\mathbf {\mathcal G}}(\I \omega_n) \right].
\end{equation} 
As our second observable we investigate  the 
normal-state linear conductance  $G=\sum_\sigma G_\sigma$ with
\begin{equation} 
\label{conform} 
G_\sigma =  \frac{e^2}{\hbar} \, \frac{2 \Gamma_L
\Gamma_R}{\Gamma_L+\Gamma_R} \, \int_{- \infty}^{\infty} A_\sigma(\omega) 
\left(-\frac{{\rm d} f(\omega)}{{\rm d} \omega}\right) \, \D \omega, 
\end{equation} 
where $A_\sigma$ denotes the normal-state dot spectral function and
$f$ the Fermi function. The computation of $A_\sigma$
from the (normal-state) imaginary frequency Green function 
${\mathcal G}_\sigma(\mathrm{i} \omega_n)$ obtained numerically by CT-INT  
is based on analytical continuation. It is found that the calculation of $G_\sigma$ is much more
reliable if the method detailed in 
Ref.~\cite{Karrasch10} is used.
As shown there the conductance can be written as  
\begin{equation} 
\label{eq:conductance_grid} 
G_\sigma = \frac{e^2}{\hbar}\frac{2
\Gamma_L \Gamma_R}{\Gamma_L+\Gamma_R} \frac{2}{\beta} \sum_{\alpha>0} R_\alpha \Im \frac{\mathrm{d}
{\mathcal G}_\sigma(\mathrm{i} \tilde{\omega}_\alpha) }{ \mathrm{d} \tilde{\omega}_\alpha}, 
\end{equation}
where the frequency derivative of the Green function has to be evaluated at imaginary 
frequencies $\I \tilde{\omega}_\alpha$ which can differ from the Matsubara ones \footnote{The choice of the frequency grid is not unique, an optimal one 
being proposed  by H.~Monien \cite{Monien06,*Monien10}. For the problem at hand however,
the additional computational effort is minor if the slightly 
less efficient method of Ref.~\cite{Karrasch10} is used instead of the optimal version, as the 
 (real) Pad{\'e} approximant of $\Im[{\mathcal G}_\sigma(\mathrm{i} \omega_n)]$ may be evaluated at
 any frequency at virtually no cost. }
given together with the weights
$R_\alpha$ in 
Ref.~\cite{Karrasch10}. Within 
CT-INT ${\mathcal G}_\sigma$ is accessible 
only at the Matsubara frequencies. Therefore, we introduce a (real) Pad{\'e} approximant 
${\mathcal G}_{\text{P}}(\omega) = \sum_{j=0}^{M-1} a_j \omega^j {\Big /}  \sum_{j=0}^{M} 
b_j \omega^j$ of degree $(M,M+1)$ and minimize the function
\begin{eqnarray*} 
\chi^2(\{a_i\},\{b_i\}) & = & \sum_{n,m} \left\{ \mathcal{G}_\text{P}(\omega_n)
- \Im[{\mathcal G}_\sigma(\mathrm{i} \omega_{n})] \right\} \\ 
&& \times C^{-1}_{n,m} \left\{ \mathcal{G}_\text{P}(\omega_m) - 
\Im[{\mathcal G}_\sigma(\mathrm{i}
\omega_{m})] \right\}, 
\end{eqnarray*} 
where $C$ is the carefully bootstrapped estimate of
the covariance matrix of the QMC data $\Im\big[{\mathcal G}_\sigma(\mathrm{i} \omega_n)\big]$. 
The degree of the Pad\'e approximant $(M,M+1)$ is chosen such that the minimal $\chi^2$ is not 
smaller than the number of degrees of freedom to obtain a statistically consistent fit and 
is found to be surprisingly small with $M=3,\dots , 6$. The Pad\'e approximant may now be 
safely evaluated at the positions $\I \tilde{\omega}_\alpha$ and Eq.~(\ref{eq:conductance_grid}) 
can be used.

\textit{Comparison to the experiment.} In experiments the charging energy 
$U$ can be determined accurately from the height 
of the Coulomb blockade diamonds obtained by bias spectroscopy in the normal state. The same 
type of measurement in the superconducting state reveals sharp features at the gap position 
from which $\Delta$ can be extracted \cite{Jorgensen07,Eichler09}. In addition, 
$T$ and $h$ are known 
within tight bounds. The parameters which are most delicate to determine but strongly 
affect $J$ are the level width $\Gamma$ and the asymmetry $\Gamma_L/\Gamma_R$. Based on 
this insight, we proceed as follows: (i) The parameters $\Delta, U, T$
and $h$ are taken directly from the experiment. Those and the comparison 
of theoretical curves for the normal-state conductance $G(\epsilon)$ 
with the experimental ones are used for obtaining accurate estimates of 
$\Gamma$, $\Gamma_L/\Gamma_R$, and the gate conversion factor $\alpha$ which
relates the change of $\epsilon$ to a variation of the gate voltage 
$V_{\rm g}$ according to $V_{\rm g}=\alpha \epsilon$ \footnote{Often $\alpha$ is extracted from the ratio of the width and 
the height of the Coulomb blockade diamonds \cite{Jorgensen07} --- a procedure which is not 
applicable in the Kondo regime as the separation of the linear conductance peaks is less 
than $U$ \cite{Costi01}.}. (ii) For the complete parameter set determined this way, we compute
the Josephson current and compare to the measured $J_{\rm c}$. 

We focus on the most symmetric conductance double peak
presented in Fig.~4d) of Ref.~\cite{Jorgensen07}. The experimental parameter 
estimates with errors of approximately 10\% are $U \approx 3\,$meV, $\Delta \approx 0.1\,$meV, $T \approx 75\,$mK, and 
$h \approx 150\,$mT. In Fig.~\ref{fig1} a) we show 
our best fit of $G(\epsilon)$ to the experimental result from which we extract 
$\Gamma=0.27\,$meV, $\Gamma_L/\Gamma_R=9.3$, and $\alpha=0.011\,$V/meV. 
At fixed $U$ the peak separation and the peak to valley ratio are determined 
by $\Gamma$ while the overall height is given by $\Gamma_L/\Gamma_R$, as can be inferred 
from Eq.~(\ref{eq:conductance_grid}) (in ${\mathcal G}_\sigma$ only $\Gamma=\Gamma_L+\Gamma_R$ enters). %
Note that 
$\Gamma$ turns out to be significantly smaller and 
$\Gamma_L/\Gamma_R$ significantly larger than the values extracted in 
Ref.~\cite{Jorgensen07} based on the 
assumption that the dot is in the Coulomb blockade regime. However, our analysis
allowing for Kondo correlations clearly reveals that those are relevant for %
$U/\Gamma \approx 11.15$ and the Kondo scale $k_{\rm B}T_{\rm K} \approx  8\,\mu$eV. It is roughly an order of magnitude smaller than $\Gamma$ and of the order of the temperature 
($k_{\rm B}T=6.5 \, \mu$eV) as well as the Zeeman energy ($\mu_{\rm B} h=8.7 \, \mu$eV). Therefore 
neither $T$ nor $h$ can be neglected when considering the normal state; the conductance 
is characterized by a split Kondo plateau (ridge) \cite{Costi01}, not to be mistaken 
with the Coulomb blockade peaks which would be located at larger energies 
$\epsilon \approx \pm U/2 \approx \pm 1.5\,$meV. As an inset we show, for illustration, the normal-state 
spectral function %
at $\epsilon =0$ %
for the extracted 
parameters obtained from analytic continuation of CT-INT data onto the real frequency axis by the %
maximum entropy method \cite{Beach04}. The appearance of a  sharp zero energy 
resonance is a hallmark of Kondo correlations \cite{Hewson93}. 
The splitting of the Kondo resonance by the Zeeman field is too small to 
be observable on the scale of the plot (but present in the data). %

In the experiments \cite{Jorgensen07,Eichler09} $J_{\rm c}$, 
defined as the maximum of $|J(\phi)|$ over $\phi \in [0,\pi]$, is extracted from current-voltage 
characteristics of the on-chip circuits using an extension of the standard RSJ model 
\cite{Ivanchenko69}. In this analysis it is assumed that $J(\phi)$ is purely sinusoidal with 
its maximum at $\phi=\frac \pi 2$. At first glance this appears to be at odds with what 
is known theoretically for the current-phase relation of a Josephson quantum dot in the $0$-phase 
(half-sinusoidal with maximum at $\phi \to \pi$) and the transition region (jump from $J>0$ to $J<0$ at 
$T=0$, smeared out by $T>0$) \cite{Choi04,Siano04,Sellier05,Karrasch08,Meng09}. 
However, as it was shown already in Ref.~\cite{Jorgensen07} for an effective noninteracting 
model, the sizable left-right asymmetry and the finite temperatures of the experimental setups 
imply sinusoidal currents in the $0$- and %
$\pi$-phase apart from very narrow ranges around the $0$-$\pi$ transitions. This conclusion is 
confirmed by the numerically exact CT-INT in Fig.~\ref{fig2}, where we present $J(\phi)$ 
for the above given parameters at the level positions indicated by the arrows in Fig.~\ref{fig1} b) showing 
$|J(\phi=\frac \pi 2,\epsilon)|$. Apparently only for $\epsilon$ very close to the transition 
the $\phi$-position of the maximal current $|J|$ deviates observably from $\frac \pi 2$ 
and yet the maximal value is still very close to that of $|J(\phi=\frac \pi 2)|$. This gives an 
a-posteriori justification of the extraction of $J_{\rm c}$ using the extended RSJ 
model and allows us to focus on the current at $\phi=\frac \pi 2$ when comparing to the gate 
voltage dependence of the critical current, as done in Fig.~\ref{fig1} b).  
Without any additional fitting parameters we achieve excellent 
agreement in both the $0$- (to the left and right of the peaks) and the $\pi$-phase (central region with 
nearly $\epsilon$-independent $J_{\rm c}$). In addition we show $|J(\phi=\frac \pi 2,\epsilon)|$ 
obtained for the same parameters using the HF approach 
\cite{Shiba73,Karrasch08}.
Whereas in the empty dot and doubly occupied regime $|\epsilon| \gtrsim 2\,\text{meV}$
this  mean-field approximation gives decent agreement with the  
exact result (CT-INT; see also Fig.~\ref{fig3})
it apparently fails in the mixed valence regime and for half dot filling ($\epsilon \approx 0$) in which Kondo correlations are crucial.   
Important features like the smoothing of the phase transition by the finite temperature and 
the smooth crossing through zero of $J(\frac \pi 2)$ cannot even be obtained qualitatively. %
This emphasizes that Kondo correlations 
are relevant even in the presence of prevailing superconductivity ($\Delta \approx 10 
T_{\rm K}$)
\footnote{Besides the failure of HF to quantitatively describe 
$J(\phi=\frac \pi 2,\epsilon)$ it suffers from the severe artifact that the $\pi$ phase is 
produced by a spurious spin-symmetry breaking. We furthermore emphasize that the HF 
normal-state linear conductance does not capture Kondo physics and thus  cannot be used to extract 
$\Gamma$ and $\Gamma_L/\Gamma_R$ from the measured $G(\epsilon)$ as is done here using CT-INT.}.  

\textit{Increasing $T_{\rm K}$.} Considering $\Gamma=0.4\,$meV and $0.5\,$meV with all the other 
parameters fixed at the values given above, we finally investigate the regime $\Delta 
\approx k_{\rm B} T_{\rm K}$ of the strongest competition between superconductivity and (precursors of) 
Kondo correlations. In  Fig.~\ \ref{fig3} $|J(\phi=\frac \pi 2,\epsilon)|$ obtained by CT-INT 
is compared to HF results. Obviously, the singlet ($0$) phase is stabilized by the 
correlations --- an effect which HF is unable to describe. For the largest $T_{\rm K}$ 
(at $\Gamma=0.5$) $|J(\phi=\frac \pi 2,\epsilon)|$ computed by CT-INT only shows a 
precursor of the $\pi$-phase (the slight suppression close to 
$\epsilon=0$) while HF gives a spurious $\pi$-phase. It would be very 
interesting to measure the gate voltage dependence of the critical current for 
dots falling into this parameter regime, which would confirm the predictive power of our
calculations.

\textit{Summary.} We presented a thorough study of the Josephson-current $0$-$\pi$ transition 
of a quantum dot in the Kondo regime. A quantitative agreement to the measured dependence of 
the critical current on the gate voltage for a dot with Kondo correlations but prevailing 
superconductivity was achieved. This shows that our minimal model 
is sufficient to quantitatively capture the relevant physics and qualifies the CT-INT as a theoretical tool with predictive power for %
transport properties of correlated quantum dots. We further studied the regime of the strongest competition between 
superconductivity and Kondo correlations confirming qualitatively that the latter stabilize the
singlet state and thus %
the $0$-phase and predicting quantitatively the supercurrent, which can be experimentally verified.  

\textit{Acknowledgments}. We are grateful to the authors of Refs.~\cite{Jorgensen07} 
and \cite{Eichler09}, in particular  K.~Grove-Rasmussen, R.~Deblock, and A.~Eichler for 
discussions.  This work was supported by the DFG  via FOR1346 (DL and FA) and KA3360-1/1 (CK) and
by the Czech Science Foundation via  grant No.~204/11/J042 (TN).

\bibliography{JosephsonKondo}

\end{document}